\documentclass[preprintnumbers,amsmath,amssymbm,prd]{revtex4}
\usepackage{epsfig}
\usepackage{graphicx}
\usepackage{amssymb}

\begin{document}
\title{Comment on ``Charged scalar field at future null infinity via nonlinear hyperboloidal evolution'' 
[Phys. Rev. D {\bf 112}, 104053 (2025), arXiv:2506.15311]}
\author{Shahar Hod}
\affiliation{The Ruppin Academic Center, Emeq Hefer 40250, Israel}
\affiliation{ }
\affiliation{The Jerusalem Multidisciplinary Institute, Jerusalem 91010, Israel}
\date{\today}

\begin{abstract}
\ \ \ The asymptotically decaying tails that characterize the late-time dynamics of collapsing 
self-gravitating charged massless scalar fields were studied three decades ago 
by Hod and Piran (HP). In particular, it was shown, both analytically and numerically, that the 
late-time behavior of these collapsing charged massless scalar fields is governed by oscillatory inverse power law tails, 
which decay more slowly than the familiar tails of neutral massless fields. 
Recently Álvares and Vaño-Viñuales (AVV) have investigated the same model numerically. 
While most of their results are in very good agreement with the earlier findings of HP, there are also some discrepancies between the original results of HP and those reported by AVV.
In this compact comment, we wish to highlight a number of inaccurate claims and critical errors 
in the analysis and results presented by AVV.
\end{abstract}
\bigskip
\maketitle


The late-time dynamics of collapsing nearly spherically symmetric neutral field configurations was first studied 
analytically in the seminal work of Price \cite{Pri} (see also \cite{Bic}). 
Interestingly, it has been revealed that, at asymptotically late times, linearized massless fields are characterized by an 
inverse power law functional behavior of the form
\begin{equation}\label{Eq1}
\psi\sim t^{-(2l+3)}\  ,
\end{equation}
where $l$ is the angular harmonic index of the neutral field mode. 

The late-time dynamics of a nearly spherically symmetric {\it charged} gravitational collapse 
was first studied by Hod and Piran \cite{HP1,HP2,HP3}. 
Intriguingly, it has been proved analytically \cite{HP1,HP2} and confirmed numerically \cite{HP3} 
that the asymptotic functional 
behavior of collapsing charged massless scalar fields is characterized by oscillatory inverse power law tails 
of the forms \cite{Noteunit}
\begin{equation}\label{Eq2}
\psi\sim e^{i{{qQ}\over{r_+}}t}\cdot t^{-(2\beta+2)}\ \ \ \ \text{at timelike infinity}\
\end{equation}
and
\begin{equation}\label{Eq3}
\psi\sim e^{i{{qQ}\over{r_+}}t}\cdot u^{-(\beta+1-iqQ)}\ \ \ \ \text{at future null infinity}\  ,
\end{equation}
where $\{Q,r_+\}$ are respectively the electric charge and the outer horizon radius of the 
charged Reissner–Nordström (RN) black hole, $q$ is 
the charge coupling parameter of the field, $u$ is the retarded time coordinate, 
and the dimensionless parameter $\beta$ is given by the functional relation
\begin{equation}\label{Eq4}
\beta(l,qQ)={{-1+\sqrt{(2l+1)^2-4(qQ)^2}}\over{2}}\  .
\end{equation} 
As explicitly emphasized in \cite{HP1,HP2,HP3}, the functional expressions (\ref{Eq1}), (\ref{Eq2}), and (\ref{Eq4}) imply 
that, at asymptotically late times, charged massless fields decay slower than neutral massless fields.   

Recently, Álvares and Vaño-Viñuales (AVV) \cite{AVV} performed 
a numerical study of the same model problem. While most of their results are in good agreement with the earlier 
findings of Hod and Piran (HP) \cite{HP1,HP2,HP3}, some discrepancies do exist. 

In what follows, we point out a number of inaccurate claims and critical errors that appear in 
the analysis and results presented in \cite{AVV}:

(1) The abstract of \cite{AVV} begins with the statement: 
``Quasinormal modes and power-law late-time decay tails of a charged scalar field in a charged black hole
background have been studied, but never in the fully nonlinear regime, as far as we know.''. 
\newline
This opening claim of \cite{AVV} is erroneous. 

In particular, it should be emphasized that the fully non-linear collapse of self-gravitating 
charged massless fields was studied numerically in \cite{HP3}, where it was explicitly shown that 
the agreement between the analytical predictions (\ref{Eq2}), (\ref{Eq3}), and (\ref{Eq4}) 
of the linearized analysis \cite{HP1,HP2} and the non-linear numerical 
results \cite{HP3} is remarkably good. 

(2) It is claimed in the introduction section of \cite{AVV} that ``...Ref. [8] extended the late-time power laws
to the charged scalar field collapsing into a RN black hole.'' (Note that Ref. [8] of \cite{AVV} is Ref. \cite{Bic} of our paper). 
\newline
This statement of \cite{AVV} is also erroneous. 

In particular, the interesting work of Bičák \cite{Bic} (Ref. [8] of \cite{AVV}) analyzed 
the late-time evolution of a {\it neutral} scalar 
field in the background of a charged RN black hole. 
As emphasized above, the late-time dynamics of a {\it charged} scalar 
field in the background of a charged RN black hole was first studied almost two decades later in \cite{HP1,HP2,HP3}. 

(3) It is claimed in the introduction section of \cite{AVV} that the late-time behavior of neutral fields at timelike infinity 
is dominated by the functional behavior $\psi\sim t^{-(2l+2)}$ [see equation (1) of \cite{AVV}]. 
\newline
This statement of \cite{AVV} is not accurate. 

In particular, it is important to emphasize that the late-time behavior $\psi\sim t^{-(2l+2)}$ 
holds only when the field is initially static (that is, 
if an $l$-pole moment is present before the collapse). 
In the general case, the late-time asymptotic behavior of the fields is described by the 
functional relation [see Eq. (\ref{Eq1})] $\psi\sim t^{-(2l+3)}$ \cite{Pri}.

(4) It is claimed in the introduction section of \cite{AVV} that the late-time behavior of charged massless 
fields at timelike infinity is dominated by an inverse power law decaying tail with exponent $-2\beta+2$ [see equation (2) of \cite{AVV}]. 
\newline
This statement of \cite{AVV} is erroneous.

In particular, it should be emphasized that, as explicitly proved in \cite{HP2}, 
the correct power-law exponent characterizing the late-time asymptotic behavior of charged massless fields 
at timelike infinity is $-(2\beta+2)$ [see Eqs. (\ref{Eq2}) and (\ref{Eq4})].

(5) We also point out that there is a similar error in equation (32) of \cite{AVV}, where it is claimed that 
the power-law exponent characterizing the late-time asymptotic behavior of charged massless fields 
at timelike infinity is $\Re(-2\beta)+2$. 
\newline
However, as explicitly proved in \cite{HP2}, the correct real part of the power-law exponent 
governing the late-time decay of charged massless fields at timelike infinity is $\Re(-2\beta)-2$. 

(6) It is interesting to note that the numerical results presented in Figs. 5-7 of \cite{AVV} for the values 
of the charge-dependent damping exponents show that the analytical 
predictions (\ref{Eq2}), (\ref{Eq3}), and (\ref{Eq4}) of \cite{HP1,HP2}, 
which are based on a linearized analysis of the field equations, are in good agreement with the corresponding 
damping exponents of fully non-linear charged fields, except in the dimensionless small-charge 
regime $qQ\lesssim10^{-3}\ll1$.

It is important to emphasize that the results presented in \cite{HP1,HP2} imply that, in the gravitational collapse 
of charged massless scalar fields with $qQ\ll1$, 
the ratio between the mass-dependent (curvature) contribution $\psi_{M}$ to the 
late-time tail and the charge-dependent (electromagnetic) contribution $\psi_{Q}$ to the late-time 
tail is given by the dimensionless relation \cite{Notetn}
\begin{equation}\label{Eq5}
{{\psi_{M}}\over{\psi_{Q}}}=O\Big({{M}\over{qQt}}\Big)\  ,
\end{equation}
where $\{M,Q\}$ are respectively the mass and the electric charge of the matter configuration. 

From the characteristic relation (\ref{Eq5}) one deduces that the late-time behavior of charged massless scalar fields 
is dominated by its charge-dependent (electromagnetic) part, as given by Eqs. (\ref{Eq2}), (\ref{Eq3}), and (\ref{Eq4}), 
in the asymptotic late-time regime 
\begin{equation}\label{Eq6}
t\gg {{M}\over{qQ}}\  ,
\end{equation}
a regime that was {\it not} explored numerically in \cite{AVV} for the case $qQ\lesssim10^{-3}\ll1$. 

In particular, as noted in \cite{AVV}, the numerically computed damping exponents of the charged fields 
presented in Fig. 7 of \cite{AVV} are extracted from the time interval $t/M\in[100,150]$ of the numerical simulations 
which, for the case $qQ\lesssim10^{-3}$, should be regarded as an intermediate (and {\it not} asymptotic) 
time regime [see Eq. (\ref{Eq6})]. 

(7) It is claimed in \cite{AVV} that 
the oscillation frequency of charged massless scalar fields at null infinity is given by the 
simple expression $qQ/10$ (see, in particular, Fig. 8 of \cite{AVV}). 

We note, however, that it is clear that the expression $qQ/10$ cannot represent a physical frequency. 
In particular, the dimensionless quantity $qQ$ (see \cite{Noteunit}) does not have the appropriate 
units of frequency.

(8) It is claimed in \cite{AVV} [see, in particular, equation (4) of \cite{AVV} and the black curve 
in Fig. 8 of \cite{AVV}] that, at null infinity, 
the oscillation frequency of the charged massless scalar fields predicted by \cite{HP1,HP2,HP3}
is given by the value $qQ-\Im\beta$  . 
\newline
This claim of \cite{AVV} is erroneous.

In particular, the partial expression (4) of \cite{AVV} incorrectly suggests 
that the scalar field is a periodic function of $\ln u$ at null infinity. 
However, an examination of Fig. 1 of \cite{HP3} and Fig. 5 of \cite{AVV} clearly shows that this is not the case.
 
In fact, the correct leading-order oscillation frequency predicted by \cite{HP1,HP2,HP3} for the charged massless 
scalar fields (as a function of the null coordinate $u$) at null infinity is given by the expression $qQ/2r_+$, 
an expression that has the correct units of frequency. 
It appears that the authors of \cite{AVV} have forgotten to include in their equation (4) 
the oscillatory factor $e^{i{{qQ}\over{r_+}}t}$ of the charged scalar field at 
future null infinity [see Eq. (\ref{Eq3})] \cite{Noteeti}.      
 
In particular, taking cognizance of Eq. (\ref{Eq3}) and using the relation $t=(u+v)/2$ \cite{Notetuv}, one finds 
the $u$-dependent oscillating (and decaying) functional behavior 
\begin{equation}\label{Eq7}
\psi\sim e^{i{{qQ}\over{2r_+}}u}\cdot e^{i(qQ-\Im\beta)\cdot\ln u}\times u^{-(1+\Re\beta)}\
\end{equation}
for the charged massless scalar fields at null infinity. 
It is worth emphasizing again that this analytically derived \cite{HP1,HP2} asymptotic functional behavior 
of the charged massless fields was confirmed in \cite{HP3} using 
direct (linear as well as non-linear) numerical computations. 

\bigskip
\noindent
{\bf ACKNOWLEDGMENTS}
\bigskip

This research is supported by the Carmel Science Foundation. I would
like to thank Yael Oren, Arbel M. Ongo, Ayelet B. Lata, and Alona B.
Tea for helpful discussions.



\begin{thebibliography}{99}
\bibitem{Noteunit} We use natural units in which $G=c=\hbar=1$. 
Note that in these units the mass $M$ and the electric charge $Q$ of the black hole have the dimensions of 
length and the charge coupling parameter $q$ of the field has the dimensions of length$^{-1}$.
\bibitem{Notetn} Note that the characteristic dimensionless ratio (\ref{Eq5}) is valid for the late-time tails 
both at timelike infinity and at future null infinity.
\bibitem{Noteeti} Likewise, it appears that the authors of \cite{AVV} have forgotten to include 
in their equation (2) the $t$-dependent oscillatory factor $e^{i{{qQ}\over{r_+}}t}$ 
of the charged scalar field at timelike infinity [see Eq. (\ref{Eq2})].
\bibitem{Notetuv} Here $u=t-y$ is the retarded time coordinate and $v=t+y$ is the advanced
time coordinate, where the ``tortoise'' radial coordinate $y$ is defined by the differential 
relation $dy/dr=(1-2M/r+Q^2/r^2)^{-1}$.
\end{thebibliography}
\end{document}